\documentclass[12pt, a4paper]{article}

\usepackage{amsmath}
\usepackage{amsfonts}
\usepackage{amssymb}
\usepackage{graphicx, rotating}
\usepackage{epstopdf}
\usepackage{epsfig}
\usepackage{latexsym}
\usepackage{graphicx}
\usepackage{color}
\usepackage{amsmath,bm,amssymb}
\usepackage{cite}
\usepackage{soul}
\usepackage{slashed}
\usepackage{float}
\usepackage{hyperref}
\hypersetup{colorlinks, citecolor=bluscuro, linkcolor=black, urlcolor=bluscuro}
\definecolor{rossos}{cmyk}{0,1,1,0.55}
\definecolor{bluscuro}{rgb}{0.15, 0.2, .85}
\definecolor{bluchiaro}{cmyk}{1,.3,0.,0.1}
\definecolor{brown}{rgb}{0.6, 0.14, 0.14}

\newcommand{\be}{\begin{equation}}
\newcommand{\ee}{\end{equation}}
\newcommand{\bea}{\begin{eqnarray}}
\newcommand{\eea}{\end{eqnarray}}

\def\CS{\cal S}
\def\CW{\cal W}
\def\CV{\cal V}

\begin{document}

\begin{titlepage}
\begin{flushright}
IFT-UAM/CSIC-18-009
\end{flushright}
\vspace{.3in}

\vspace{1cm}
\begin{center}
{\Large\bf\color{black} Terminal Holographic Complexity}\\

\bigskip\color{black}
\vspace{1cm}{
{\large J.~L.~F. Barb\'on and  J.~Mart\'{\i}n-Garc\'{\i}a }
\vspace{0.3cm}
} \\[7mm]
{\it {Instituto de F\'{\i}sica Te\'orica,  IFT-UAM/CSIC}}\\
{\it {C/ Nicol\'as Cabrera 13, Universidad Aut\'onoma de Madrid, 28049 Madrid, Spain}}\\
{\it E-mail:} \href{mailto:jose.barbon@uam.es}{\nolinkurl{jose.barbon@uam.es}}, \href{mailto:javier.martingarcia1@gmail.com}{\nolinkurl{javier.martingarcia1@gmail.com}}\\
\end{center}
\bigskip

\vspace{.4cm}

\begin{abstract}
We introduce a quasilocal version of holographic complexity adapted to `terminal states' such as spacelike
singularities. We use a modification of the action-complexity ansatz, restricted to the past domain of dependence of the terminal set, and  study a number of examples whose symmetry permits explicit evaluation, to conclude that this quantity enjoys
monotonicity properties after the addition of  appropriate counterterms. A notion of  `complexity density' can be defined for singularities  by a coarse-graining procedure. This definition assigns 
finite complexity density to black hole singularities but vanishing complexity density to either generic FRW singularities or chaotic BKL singularities. We comment on the similarities and differences with Penrose's Weyl curvature criterion. 

\end{abstract}
\bigskip

\end{titlepage}


\section{Introduction}

\noindent

Spacetime singularities are perhaps the most radical  boundary of knowledge in theoretical physics, a condition greatly amplified by the  high degree of self-consistency of General Relativity. A number of timelike singularities have been successfully resolved in string theory, through the expedient of exhibiting extra light degrees of freedom localized at the  singular locus. 
Spacelike singularities, on the other hand, are usually regarded as intrinsically associated with strong gravitational dynamics and thus  beyond the realm of string  perturbation theory.  

Very broadly, there are  two traditions regarding the interpretation of spacelike singularities: either they must be `resolved' so as to restore some type of  evolution across the singularity, or they must be accepted as  true `spacetime terminals'. To the extent that the black hole singularity is a general guide, the second option is preferred in modern discussions based on holography. On the other hand, the straightforward application of holographic ideas requires an identification of appropriate AdS/CFT boundaries or at least some notion of holographic screen. 

Recently, key roles for quantum entanglement and quantum complexity in the workings of holography have been increasingly appreciated. Roughly, the degree of connectivity of space is related to the entanglement of the holographic degrees of freedom, and
its volume behaves as a measure of their quantum complexity. In this note we seek to associate holographic measures of quantum complexity to states  which are linked to spacelike singularities by time evolution. This program was initiated in previous work \cite{BarbonRabinoComplexity, rabinonew}
by the analysis of certain cosmological singularities with controlled AdS/CFT embedding. Here we seek to provide quasilocal notions of complexity which may be abstracted from particular AdS/CFT constructions, and  therefore   having a larger degree of generality. While we use the volume-complexity (VC) proposal \cite{SusskindEntnotEnough} as a heuristic guide, most of our discussion is tailored to the more covariant action-complexity (AC) proposal \cite{SusskindCAcorto, BrownSusskindAction}.

The connection between spacetime singularities and complexity has a long history, going back to the occurrence of classical chaos in generic cosmological singularities   \cite{misner, BKL,BKL2,BKL3} (see \cite{libro} for  a recent review.) 
  In \cite{Penrose} Penrose gave a local criterion for the complexity of a singularity. The basic observation is that `ordered' singularities, such as those arising in FRW models,
have vanishing Weyl curvature, whereas more generic ones, such as those arising in gravitational collapse, have a generically divergent Weyl tensor.  Penrose argued that the Weyl criterion would be associated to a large gravitational entropy flowing into the singularity, a suggestion based on the heuristic picture of a generic cosmological crunch, full of chaotic black hole collisions. Since black holes are known to carry entropy, a corresponding notion of entropy may be assigned to the union of all singularities enclosed by the colliding black holes. 

One basic observation of this paper is that  a suitable version of holographic complexity, rather than entropy, provides a more natural measure  of complexity of a singularity. The proposal uses a restriction of  the standard AC ansatz of \cite{SusskindCAcorto, BrownSusskindAction} to the causal domain of dependence of the singularity. Furthermore, we will see that a local notion of complexity, different from Penrose's Weyl curvature, can be naturally introduced in the holographic formalism through a coarse-graining procedure. 

The paper is organized as follows. In section \ref{sec:quasilocal} we lay  down some general definitions  of `terminal holographic complexity'. In section \ref{sec:local} we discuss the local contribution to complexity and the coarse-graining procedure. In section \ref{sec:mono} we investigate the monotonicity properties of this quantity in some  examples of singularities admitting a completely analytic treatment.  Finally, in section \ref{sec:conclusions} we offer some conclusions and outlook for plausible generalizations.

\section{A quasilocal AC ansatz for terminals}
\label{sec:quasilocal}
\noindent

A rather intuitive notion of holographic complexity is provided by the VC ansatz of \cite{SusskindEntnotEnough}. Formally, it looks like  a generalization of the HRT construction for holographic entanglement entropy \cite{rt, hrt}, removing one unit of codimension. One considers extremal codimension-one surfaces anchored on boundary holographic data, and their volume, in an appropriate normalization, yields the VC complexity of the dual state (see also \cite{Mohsen}.) A more covariant prescription (AC complexity) was subsequently introduced in  \cite{SusskindCAcorto, BrownSusskindAction}, where one is instructed to integrate the bulk classical action over the full causal domain of dependence of the extremal surfaces, henceforth referred as the Wheeler-de Witt  (WdW) patch.

The basic physical guide is that either ansatz provides a linear growth of quantum complexity for a high-temperature CFT thermofield double or, in the dual picture, a large eternal AdS black hole,
 \be\label{crate}
  {dC \over dt} \sim T\,S \sim M\;,
 \ee
where $S, T, M$ denote entropy, temperature and mass, respectively. This law is supposed to apply for $t  \gg T^{-1}$, up to $O(1)$ coefficients. The detailed dependence on these coefficients is argued to be more uniform for the  AC prescription, although the
 physics is qualitatively the same, at least for large enough temperatures.\footnote{It has been conjectured that (\ref{crate}) should saturate the Lloyd bound \cite{lloyd}. See, however \cite{montero, Myerstdep}.} Still, some qualitative differences in the AC/VC dichotomy persist,  particularly for cold systems, such as near-extremal black holes or cold hyperbolic black holes. This is testimony of our still quite  poor understanding of the duality \cite{BarbonMartinHyperbolic, MyersFormation, noncomp}. 

 \begin{figure}[t]
$$\includegraphics[width=10cm]{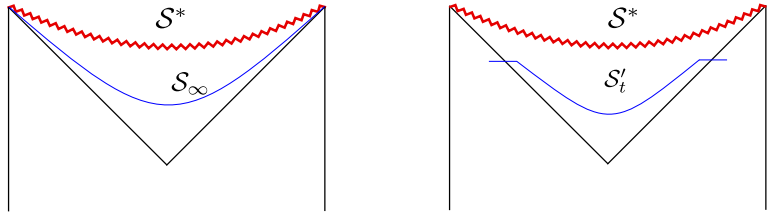} $$   
\begin{center}
\caption{\emph{  On the left, the codimension-one asymptotic surface ${\cal S}_\infty$, accounting for the total complexity `flowing' into the black-hole singularity $\CS^*$. On the right, the subtracted codimension-one surface ${\CS}'_t$ which accounts for the time-dependence of VC complexity in the eternal black hole geometry. }\label{fig:sigmainf}}
 \end{center}
\end{figure}

In the benchmark model provided by the eternal black hole spacetime, the central object of 
interest for the VC ansatz is the extremal codimension-one surface $\CS_\infty$ shown in Figure \ref{fig:sigmainf}. This surface maximizes
the volume locally and it lies entirely within the black hole interior, i.e. the past causal domain of the singularity.

The growth of complexity within the VC ansatz can be seen as the result of gradually accessing an increasing portion of $\CS_\infty$. More precisely, the portion of the extremal surface ${\CS}_t$ which has a significant contribution to time dependence  can be analyzed approximately as composed of two parts: a subset of $\CS_\infty$ with volume proportional to $t$, and a transition surface at the horizon, whose contribution to the complexity is of order $S$, the entropy of the black hole. Let us denote by ${\CS}'_t$ this, loosely defined,  `subtracted' surface as indicated in Figure \ref{fig:sigmainf}.

\begin{figure}[h]
$$\includegraphics[width=10cm]{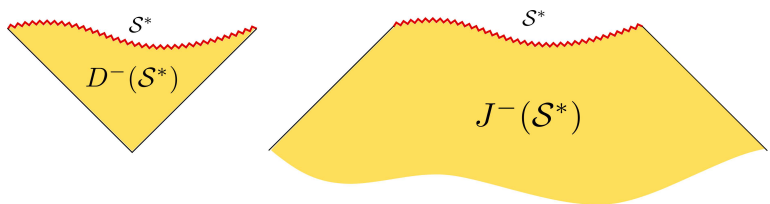} $$   
\begin{center}
\caption{\emph{Generic terminal set $\CS^*$ (in red) and its past domain of dependence $D^-(\CS^*)$ and causal past $J^-(\CS^*)$ .  }\label{fig:DS}}
 \end{center}
\end{figure}

Once we decide to focus on ${\cal S}'_t$ and its asymptotic limit $ {\cal S}_\infty$, we may consider  versions of these quantities for any terminal set $\CS^*$ (which may in particular be a proper  subset of a wider one.) The reason is that the analog of ${\cal S}_\infty$ always exists given any spacelike terminal set $\CS^*$ and its associated past domain of dependence $D^-(\CS^*)$  (see Figure \ref{fig:DS}.) Since the volume is positive and the past boundary of $D^- (\CS^*)$ is null, the extremal surface is either a local maximum of volume or it coincides with $\CS^*$ in a degenerate case. The first situation occurs when $\CS^*$ is a standard singularity of the kind we encounter at black holes and cosmological crunches in General Relativity, since the volume of spatial slices vanishes at such singularities.

\begin{figure}[h]
$$\includegraphics[width=5cm]{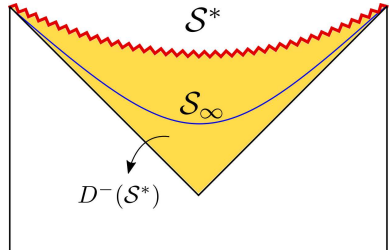} $$   
\begin{center}
\caption{\emph{  The total VC complexity flowing into the singular set ${\cal S}^*$ is the volume of the asymptotic surface ${\cal S}_\infty$. Its AC analog is the on-shell action integrated over the past domain of dependence $D^- ({\cal S}^*)$.  }\label{fig:c1}}
 \end{center}
\end{figure}

A more covariant version of $\CS_\infty$ and ${\CS}'_t $ could be obtained by adapting the AC ansatz to this situation (cf. Figure \ref{fig:c1}.) Since ${\cal S}_\infty$ is the extremal surface on $D^-({\cal S}^*)$, the natural AC version of the full  terminal complexity of the set ${\cal S}^*$ is the on-shell action 
\be\label{fullc}
C[{\cal S}^*]  \propto I\left[D^-({\cal S}^*)\right]
\;,
\ee
evaluated over the set $D^- ({\cal S}^*)$. Since this definition  only makes reference to the terminal set ${\cal S}^*$ we regard this notion of complexity as `quasilocal' and will often denote it as such. 

 \begin{figure}[h]
$$\includegraphics[width=5cm]{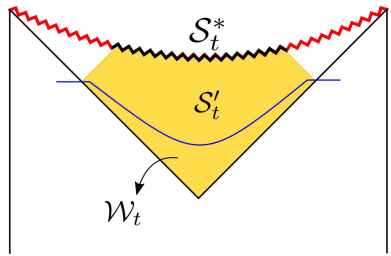} $$   
\begin{center}
\caption{\emph{  The WdW patch ${\cal W}_t$, associated to the cut-off surface ${\cal S}'_t$, intersects the singularity at ${\cal S}^*_t$. }\label{fig:c2}}
 \end{center}
\end{figure}

Next, a  notion of `time-dependence'  can be defined by considering a WdW patch anchored roughly at the exit points of the cut-off surface ${\cal S}'_t$, as indicated in  Figure \ref{fig:c2}. A more invariant definition can be obtained by noticing that these WdW patches are
nested into one another as time increases. For sufficiently `late' WdW patches, this `nesting' extends to the intersections of the WdW patches with the singular set. This suggests that  we may use the nested singular subsets as a starting point in the definition of the WdW nested family. To be more precise, let us pick  a sequence of terminal subsets ${\CS}^*_u$, ordered by inclusion in the sense that
\be
{\CS}^*_u \subset {\CS}^*_{u'}  \;, \; {\rm for}\;\; u<u'\;,
\ee
and  converging to 
the full terminal set $\CS^*$ as $u\rightarrow u_*$, we can consider a set of WdW patches ${\cal W}_u$, defined as the intersection between $D^- (\CS^*)$ and 
the causal past of ${\CS}^*_u$, 
\be\label{wdwu}
{\cal W}_u = J^- ({\CS}^*_u) \cap D^- ({\CS}^*)\;.
\ee

\begin{figure}[h]
$$\includegraphics[width=8cm]{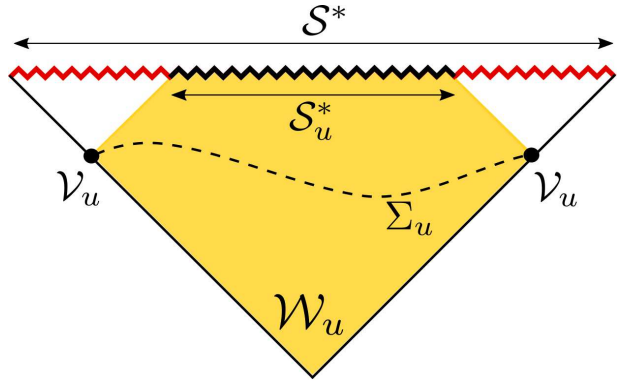} $$   
\begin{center}
\caption{\emph{ The WdW patch ${\CW}_u$ (in yellow), associated to a given ${\CS}^*_u$ subset (in black) of the full terminal set $\CS^*$ (in red).  The codimension-two set ${\CV}_u$ is the (possibly disconnected) boundary of Cauchy surfaces $\Sigma_u$ for ${\CW}_u$. }\label{fig:wdwu}}
 \end{center}
\end{figure}

For any given ${\cal W}_u$, its Cauchy surfaces $\Sigma_u$  have a common codimension-two boundary ${\cal V}_u = \partial \Sigma_u$ (cf. Figure \ref{fig:wdwu},) which would hold the `holographic data' for ${\cal W}_u$. For example, ${\cal V}_u$ is a spatial section of the event horizon when $\CS^*$ is a black-hole singularity.   Therefore, we would like to interpret the `area' of ${\CV}_u$ in Planck units \footnote{ We henceforth refer to  codimension-two volumes as `areas'.} as a measure of the effective number of holographic degrees of freedom `flowing' into the terminal subset ${\CS}^*_u$. 

In defining the WdW patches ${\cal W}_u$ we may give privilege to the `anchors', namely the codimension-two sets ${\cal V}_u$, or alternatively we may consider  the nested family ${\cal S}^*_u$, as more fundamental. These two constructions  are not completely equivalent, since the WdW patch anchored at ${\cal V}_u$ may fail to intersect ${\cal S}^*$ at sufficiently `early times'. In this paper
we are more interested in the asymptotic `late-time' behavior in which ${\cal W}_u$ does have a non-trivial boundary component at the singularity. Therefore, we tacitly adopt in what follows the nesting construction of the WdW patches and we will often refer to the
associated complexity measures as `nesting complexity'.

In the AC/VC heuristic correspondence, the codimension-two surfaces ${\CV}_u$ are the natural analogs of the transition surface with volume of order $S$ in Figure 1. Therefore, interpreting the $u$ coordinate as a (null) time variable, we are led to the following definition
of nesting complexity associated to the given family of WdW patches ${\cal W}_u$, 
\be\label{defc}
C^*_u= \alpha \,I[{\CW}_u]  + {\lambda \over 4G} \,{\rm Area}[{\CV}_u]
\;,
\ee
where $\alpha$ is a normalization factor,   $\lambda$ is an undetermined constant which sets the relative importance of the codimension-two boundary counterterm, and  
  $I[{\CW}_u]$ denotes the on-shell gravitational action, now integrated over the WdW patch ${\cal W}_u$. The action can be written as $I=I_{\rm bulk} + I_\partial$, separating bulk and boundary contributions. The bulk term has the standard form 
\be\label{wdwon}
I[{\CW}_u]_{\rm bulk} =  {1 \over 16\pi G} \int_{{\CW}_u} \left(R-2\Lambda + {\cal L}_m \right)  + \dots  \;,
\ee
 where the dots stand for higher-derivative corrections and ${\cal L}_m$ is the Lagrangian density for non-gravitational degrees of freedom, out of which we have explicitly singled out the cosmological constant. 
 
 The boundary term $I_\partial$  requires special consideration. On general grounds, it is given by a sum of contributions from  codimension-one and codimension-two components of the boundary $\partial {\cal W}_u$. The non-null codimension-one pieces and their joints  are given by the standard York--Gibbons--Hawking (YGH) term and a set of well understood joint contributions (see \cite{Poisson} for a review.) On the other hand, some formal choices are necessary in the  presence of null codimension-one pieces, and the physics behind these choices remains somewhat unclear (see for example the considerations in \cite{Poisson, Ross, MyersFormation}.)

 For the purposes of this paper, we make a minimal choice for $I_\partial$ in which we only retain the YGH term for non-null codimension-one components and we drop the contributions from codimension-one null components and their codimension-two joints. We do this while keeping open the possibility that a further understanding of the microscopic definition of complexity will require the specification of non-geometrical quantities in the AC rules.

Applying these rules to the WdW patches (\ref{wdwu}) we have an action 
\be\label{ac}
I[{\cal W}_u] = I[{\cal W}_u]_{\rm bulk} + I[{\cal S}_u^*]_{\rm YGH}\;,
\ee
where the bulk term is given by (\ref{wdwon}) and the YGH term is restricted to the spacelike singular component of the WdW patch, 
\be\label{yghs}
I[{\cal S}^*_u]_{\rm YGH} = {1\over 8\pi G}\int_{{\cal S}^*_u} K 
\;.
\ee

Hence, the unpacked  ansatz for the nesting complexity reads 
\be\label{coma}
C^*_u =\alpha  I[{\cal W}_u]_{\rm bulk} + \alpha I[{\cal S}^*_u]_{\rm YGH} + {\lambda \over 4G} {\rm Area}\,[{\cal V}_u]\;.
\ee

 Although we have chosen to regard the entropy counterterm as separate from $I_\partial$, we may as well consider it as one more boundary contribution to the action.\footnote{Incidentally, this  would correspond to a very special case of the prescription  introduced in \cite{Poisson}, in which one drops the codimiension-one null pieces and adjusts  the normalization conventions of affine parameters in an {\it ad hoc}, $u$-dependent manner.}  In this case we are effectively picking out boundary components supported on  ${\CS}^*_u \cup {\CV}_u$, rather than the full $\partial {\cal W}_u$. This instruction   admits a nice topological interpretation, namely it amounts to focusing  on the intersection of the boundaries of $J^- ({\CS}^*_u) $ and $D^- (\CS^*)$, rather than the boundary of the intersection:
\be\label{bba}
{\cal S}_u^* \cup {\cal V}_u  =\partial J^- ({\CS}^*_u) \cap \partial D^- (\CS^*)  \;,
\ee
a relation that may be used to provide an invariant definition of ${\CV}_u$ given the family of nested sets  ${\CS}^*_u$.

\begin{figure}[h]
$$\includegraphics[width=6cm]{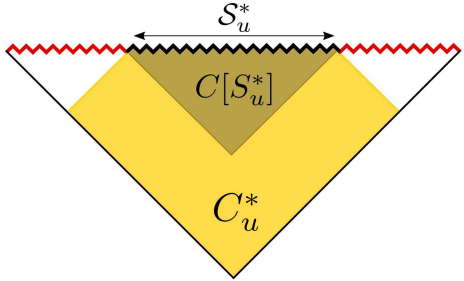} $$   
\begin{center}
\caption{\emph{ The difference between $C^*_u$ and $C[{\cal S}^*_u] $ as determined by the different domains of integration. }\label{fig:nest}}
 \end{center}
\end{figure}

  Once this nesting complexity is defined, we can now
recover the notion of `total complexity flow' into the singularity, which was loosely defined in (\ref{fullc}), as the asymptotic limit of the nesting procedure. More precisely, we have
\be\label{toal}
C[{\cal S}^*] =  \lim_{u\to u_*} C^*_u \;,
\ee
It is important to notice that, when considering singular subsets ${\cal S}_u^*$, the nesting complexity $C^*_u$ is {\it different} from the `total complexity'   $C[{\CS}^*_u]$ flowing into ${\cal S}^*_u$, as shown in Figure \ref{fig:nest}. In other words,  
we regard $C[\CS^*]$ as the AC-analog of ${\CS}_\infty$, (cf. Figure \ref{fig:c1},)  and $C^*_u$ as the AC-analog of ${\CS}'_t$, (cf. Figure \ref{fig:c2}.)

In the rest of the paper we shall see that the  definition of $C^*_u$ given in (\ref{coma}) satisfies two interesting properties of a general character. 
The first is the existence
of a local component defined at space-like singularities which can be reached by a coarse-graining procedure (Section \ref{sec:local},) and the second is the conjectured monotonicity of the
nesting complexity $C^*_u$, to be discussed in Section \ref{sec:mono} below.

\section{The local component of the terminal complexity}
\label{sec:local}

\noindent

A remarkable property of the AC complexity prescription is the occurrence of a quantitatively important  contribution coming directly from the singularity,  through the evaluation of the YGH term. Since this is a term in the low-energy effective action, we should be suspicious of its validity. On the other hand, we are instructed to take this contribution seriously down to its precise dependence on coefficients, as this is crucial for the claimed
uniformity of the growth law (\ref{crate}) for AdS black holes in various dimensions, large and small.   In a similar vein, the contribution (or lack of it) of the YGH term at the singularities is crucial for the `non-computing' behavior in various  systems, such as AdS black holes in the $1/d$ expansion \cite{noncomp}
 and cold hyperbolic black holes \cite{BarbonMartinHyperbolic, MyersFormation}.

These considerations suggest that complexity is actually the piece of the holographic dictionary which most efficiently `sees' the properties of the singularities. 

The YGH contribution is local and formally extensive over the singular set ${\cal S}^*$. However, the volume form is not generally defined at ${\cal S}^*$, which makes the notion of `extensivity' non-trivial.  In order to elucidate this point, let us parametrize the 
  near-terminal metric by  a Gaussian normal  coordinate $\tau$. This foliates  the near-terminal  spacetime into spacelike surfaces $\Sigma_\tau$, 
according to the proper-time distance to $\CS^*$. In defining a metric on the $\Sigma_\tau$ slices, we extract a conventional power of the proper time according to the ansatz 
\be\label{locs}
ds^2 = -d{\tau}^2 + (\tau H)^{2\gamma/d} \, d\Sigma_\tau^{\;2}\;.
\ee
Here, 
$H$ is an inverse-length setting a characteristic value for the expansion away from  the terminal set. 
In general,  the $d$-dimensional  metric  $d\Sigma_\tau^2$  does not have a smooth limit  as $\tau \rightarrow 0^+$, but we may choose the conventional exponent $\gamma$
in such a way that its volume form   does have a smooth limit. We shall actually assume that this volume form  is analytic in $\tau$, since this will be a property of all examples we study (it would be interesting to assess the generality of this assumption.) We will refer to such notion of volume for ${\CS}^*$ as the `comoving volume' of the terminal set and 
denote its measure as $d{\rm Vol}_c $.

In this notation, the YGH term in the action is  computed as
\be\label{ygh}
I[{\CS}^*]_{\rm YGH}= {1\over 8\pi G} \lim_{{\tau}\rightarrow 0^+} \partial_{\tau} \big[(H{\tau})^{\gamma} \,{\rm Vol}_c [\Sigma_\tau]\big]\;.
\ee
 Picking the term proportional to the comoving volume ${\rm Vol}_c [{\cal S}^*]$ of the singular set,  we find that the YGH term vanishes for $\gamma >1$ and is infinite for $\gamma <1$, except perhaps the case $\gamma =0$ where the answer depends on the possible occurrence of logarithmic terms in the terminal expansion near $\tau=0$. The most interesting case is $\gamma =1$, for which one defines a nontrivial  `comoving complexity density' at the singular set, given by $H/8\pi G$. 
 
 The black hole singularity has $\gamma =1$ and thus presents a purely local contribution to complexity. In fact, this feature appears to be quite general. At spherically symmetric black-hole singularities we have a vanishing ${\bf S}^{d-1}$ and an expanding `radial' direction. Hence, the metric is locally of the Kasner form, i.e. 
 \be\label{km}
 ds^2 = -d\tau^2 + \sum_{j=1}^d (H\tau)^{2p_j } d\sigma_j^2\;,
 \ee
 with a particular choice of Kasner parameters $p_c = 2/d$ for  $d-1$ `crunching' directions and $p_r = -1 + 2/d$ for the `ripping' direction. More generally, the Kasner parameters are restricted to satisfy 
 the sum rules $\sum_j p_j  = \sum_j p_j^2 =1$ and any such metric can be put in the form (\ref{locs}) with $\gamma =1$, with
 `comoving' metric
  \be\label{comov}
 d\Sigma_\tau^2 = \sum_j (\tau H)^{2p_j  - 2/d} \,d\sigma_j^2
 \;.
 \ee
  In particular, it has a smooth comoving volume form,  
 \be
 d{\rm Vol}_c [ \Sigma_\tau] = \wedge_{j=1}^d d\sigma_j
 \ee
  as a simple consequence of the sum rule $\sum_j p_j =1$.

   The $\gamma=1$ property and the resulting non-vanishing `complexity density' persist if we let the Kasner parameters depend smoothly  on the `longitudinal' $\sigma_j$. In fact, the  classic results of ref. \cite{BKL,BKL2,BKL3} (BKL) indicate that such a `generalized Kasner'  metrics 
 furnish a good local approximation of the near-singular region (after a slight generalization involving local rescalings and frame rotations.) 
 
\subsection{Evanescent terminal complexity}

\noindent

If we regard the generalized Kasner behavior as `generic' we may say that spacelike singularities tend to have a non-vanishing  local complexity density. On the other hand, there are important examples of singularities whose YGH contribution vanishes, such as those  occurring in standard FRW spacetimes. 

To bring this simple point home, we can apply (\ref{ygh}) to the standard FRW metric
\be
ds^2 = -d\tau^2 + a(\tau)^2 \,d\Sigma^2\;,
\ee
with a singularity at $\tau=0$. By construction, the non-singular comoving volume is just given by the volume of the homogeneous and isotropic surfaces $\Sigma$, so that the complexity exponent  $\gamma$ can be read off from the short-time asymptotics of the scale factor $a(\tau)$. Since FRW metrics require non-trivial matter degrees of freedom, we follow standard practice and model them as a perfect fluid with squared speed of sound equal to $w = p/\rho$, where $p$ denotes the pressure and $\rho$ the energy density. Then, we have the standard solution $\rho \;a^{d(1+w)} = $ constant, which leads to 
$a(\tau) \sim \tau^{2 /d(1+w)}$ or, equivalently 
\be
\gamma_{\rm FRW} = {2 \over 1+w}\;.
\ee
This result implies that the only FRW singularity with a finite complexity density is the slightly unphysical case with `stiff matter', $w=1$, leading to $\gamma=1$. On the other hand, a formally infinite contribution to the complexity density, associated to $\gamma <1$, would require $w>1$  in the FRW context, i.e. a violation of  the physical conditions on the matter degrees of freedom. 

Imposing the physical condition that the matter is strictly below the `stiff' limit, $w<1$, we have $\gamma >1$, implying  a {\it vanishing} local complexity. 
Hence, we find that  `ordered' singularities of FRW type have a vanishing
local contribution to holographic complexity, just as it happened with the Weyl criterion of Penrose.

It turns out that there is an interesting twist to this story. According to the classic  BKL analysis \cite{BKL,BKL2,BKL3}, the vicinity of a generic spacelike singularity is not quite described by a single generalized Kasner metric, but rather 
 an oscillating  regime where a series of `epochs' succeed one another, each epoch being locally described by a generalized Kasner solution of the type (\ref{km}). The values of the Kasner parameters, $p_j$, change from one epoch to the next in a deterministic but chaotic manner. In addition, the frame determining the special coordinates $\sigma_j$ in (\ref{km}) undergoes a rotation, and furthermore the induced volume form at fixed $\tau$
is rescaled by a finite factor which we may absorb in the dimensionful expansion parameter $H$. Hence, in the $n$-th epoch we
have a metric
\be\label{nth}
ds^2 |_{(n)} = -d\tau^2 + (H_n \tau)^{2/d} \;d^2\Sigma_\tau^{(n)}\;,
\ee
where $d^2 \Sigma_\tau^{(n)}$ is a rotated version of (\ref{comov}) with Kasner parameters $p_j^{(n)}$. All epochs are described
by $\gamma=1$ metrics but, crucially, they have slightly different parameters $p_j^{(n)}$ and $H_n$.  In particular, the substitution rule for the expansion parameter is 
\be\label{eva}
H_{n+1} = (2p_r^{(n)} + 1) \,H_n\;,
\ee
where $p_r^{(n)} <0$ is the `ripping' parameter of the $n$-th epoch. Since $2p_r^{(n)} + 1 < 1$ for all $n$, the series of $H_n$ is
monotonically decreasing.

 If we compute
the YGH contribution to complexity by placing a regulating surface and taking the limit, the result of the complexity density is determined by the limit of $H_n$. Namely it is proportional to 
\be\label{infp}
 \prod_{\rm epochs} (2p_r +1)\;.
 \ee
 According to the analysis of \cite{BKL,BKL2,BKL3} the truly generic singularity features an infinite number of Kasner epochs. In this situation  the  product (\ref{infp}), featuring  an infinite set of numbers in the open interval $(0,1)$,  is bound to vanish for almost all singularities. We refer to this phenomenon as the `evanescence' of the local complexity for a generic BKL singularity. We may argue that, ultimately, a cutoff at Planck time from the singularity must be imposed but, in any case, the complexity computed by this ansatz would have a suppression factor determined by the number of epochs taking place until Planck time. These
 arguments suggest that the generic singularity is not that different from the FRW one, and the standard black-hole singularities are the `special ones'
 regarding complexity. 

\subsection{Local terminal complexity and coarse-graining}

\noindent

The remarkable properties of the local YGH contribution  beg the question of whether we may be able to
isolate this term in more physical terms. A natural strategy in this case is to focus on the extensivity of the local contribution, a property not shared by the full AC complexity. We can illustrate this point by focusing on the simpler case of vacuum solutions.

\begin{figure}[t]
$$\includegraphics[width=8cm]{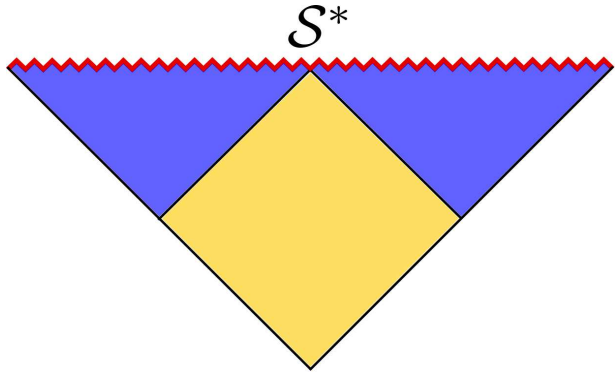} $$   
\begin{center}
\caption{\emph{ When the bulk action is dominated by a negative cosmological constant, the bulk contribution to the AC terminal complexity of  $\CS^*$ is subextensive. While the YGH contribution is extensive over $\CS^*$, the bulk contribution is more negative for the larger set (yellow) than it is for
the union of the smaller sets (blue).  }\label{fig:subextensive}}
 \end{center}
\end{figure}

\begin{figure}[t]
$$\includegraphics[width=8cm]{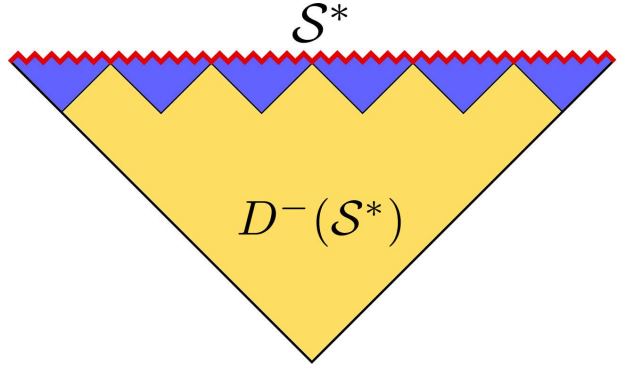} $$   
\begin{center}
\caption{\emph{ When the bulk volume remains sufficiently bounded in the vicinity of the terminal surface, the local complexity of ${\CS}^*$ results from the limit of an extreme coarse graining.  }\label{fig:ultralocal}}
 \end{center}
\end{figure}

A vacuum solution is a $(d+1)$-dimensional Einstein manifold whose metric satisfies
\be\label{eme}
R_{\mu\nu} = {2\Lambda \over d-1} g_{\mu\nu}\;,
\ee
with cosmological constant $\Lambda$ and no matter degrees of freedom. The bulk contribution to the on-shell action is then
proportional to the spacetime volume
 \be\label{bulkc}
 I[X]_{\rm bulk} = {1\over 16\pi G} \int_X (R-2\Lambda) = {\Lambda \over 4\pi G (d-1)} \,{\rm Vol} [X]\;.
 \ee
Assuming a $\gamma=1$ singular set ${\cal S}^*$  with non-vanishing complexity density, we have a full  complexity given formally by
\be\label{fullq}
C[{\cal S}^*] =  \alpha{\Lambda \over 4\pi G (d-1)} {\rm Vol}\left[D^- ({\cal S}^*)\right] + \alpha {H \over 8\pi G} {\rm Vol}_c [{\cal S}^*] \;.
\ee
While the YGH term is extensive along the comoving volume of ${\cal S}^*$, the bulk contribution is extensive in the full spacetime volume of the past domain of dependence. Considering the case $\Lambda <0$, as corresponds to states in an AdS/CFT context, we have a negative-definite bulk contribution, leading to a `subextensivity' property of the full quasilocal complexity. Indeed, under a
coarse-graining of the singular set ${\cal S}^* = \cup_i {\cal S}^*_i$ as indicated in Figure \ref{fig:subextensive}, the expression (\ref{fullq}) satisfies 
\be\label{sube}
C[{\cal S}^*] = C\left[\cup_i {\cal S}^*_i \right] \leq \sum_i C[{\cal S}^*_i]\;.
\ee
The inequality is reversed (corresponding to superextensivity) for vacuum singularities in $\Lambda >0$ spaces. The deviation from
extensivity  would disappear if the bulk contributions were to become negligible, a situation we may expect in the limit of extreme coarse graining, illustrated in Figure \ref{fig:ultralocal}. 

\begin{figure}[t]
$$\includegraphics[width=8cm]{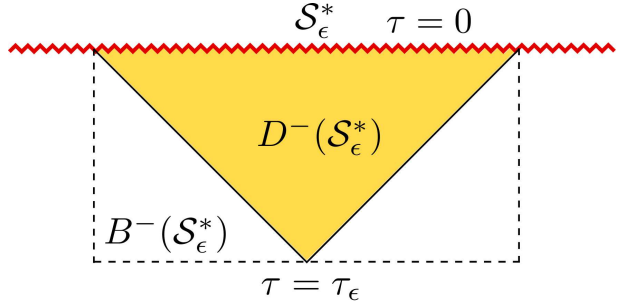} $$   
\begin{center}
\caption{\emph{ Comparison between the $\epsilon$-box and the past domain of dependence of ${\cal S}^*_\epsilon$. }\label{fig:box}}
 \end{center}
\end{figure}

In this case, the limit of infinite coarse-graining does isolate the YGH term. To be more precise, we require that the bulk contributions be consistently smaller than the YGH contribution for small subsets of ${\cal S}^*$. We can check this explicitly for $\gamma =1$ vacuum singularities  described by (\ref{fullq}) and admitting a local Kasner description. Let us consider a fine partition of the singular set by subsets ${\cal S}^*_\epsilon$ with comoving volume of order $\epsilon^d$. The condition for the bulk contribution to be negligible for small sets is that
\be\label{epd}
{{\rm Vol}\left[D^- ({\cal S}^*_\epsilon)\right] \over {\rm Vol}_c \left[{\cal S}^*_\epsilon\right]} \sim \epsilon^{\,a}
\;,\ee
with $a >0$. 
Instead of computing the volume of the past domain of dependence, $ D^- ({\cal S}^*_\epsilon)$,  it is easier to compute the volume of the larger set $B^- ({\cal S}^*_\epsilon)$, which `boxes' it
in the standard coordinate frame. If $\tau_\epsilon$ is the maximal value of the $\tau$ coordinate in $D^- ({\cal S}^*_\epsilon)$, the
$\epsilon$-box is defined by the full $\tau\leq \tau_\epsilon$ subset with given comoving coordinates covering ${\cal S}^*$, c.f. Figure (\ref{fig:box}). 
Evidently, ${\rm Vol}\left[B^-({\cal S}^*_\epsilon)\right] \geq {\rm Vol}\left[D^- ({\cal S}^*_\epsilon)\right]$, so that it is enough to establish the condition (\ref{epd}) for $B^- ({\cal S}^*_\epsilon)$.

In order to construct explicitly $B^- ({\cal S}^*_\epsilon)$  for the Kasner metric (\ref{km}) we define ${\cal S}^*_\epsilon$ to be a  $d$-dimensional cube  in the $\sigma$ coordinates with common extent $\Delta \sigma_j = \epsilon$. Its comoving volume is ${\rm Vol}_c \left[{\cal S}^*_\epsilon\right] = \epsilon^d$ and the past domain of dependence, $D^- ({\cal S}^*_\epsilon)$, is a trapezoid with base ${\cal S}^*_\epsilon$ and a  ridge with the topology of a $(d-1)$-dimensional cube, determined by the intersection of light rays in the spacetime plane with faster past-convergence. For any direction $\sigma_j$ we can define a corresponding conformal time coordinate $\eta_j$ such that light rays propagate with unit slope in the $(\eta_j, \sigma_j)$ plane. The explicit relation between $\eta_j$ and the proper time is
\be
(1-p_j) H \eta_j = (H\tau)^{1-p_j}
\;,
\ee
where $p_j$ is the Kasner exponent in the direction $\sigma_j$. Light rays whose $\sigma_j$ separation is $\epsilon$ at  $\tau=0$ converge in the past at $\tau_\epsilon^{(j)}$ given by
\be
H\tau_\epsilon^{(j)} = \left((1-p_j) {H\epsilon \over 2}\right)^{1\over 1-p_j}\;.
\ee
  Thus, the past domain of dependence of the full ${\cal S}^*_\epsilon$ set is determined by the smallest $\tau_\epsilon^{(j)}$ or, equivalently, by the {\it largest} Kasner exponent which we denote by $p_c$:
\be\label{taue}
H\tau_\epsilon = \left((1-p_c){H\epsilon \over 2}\right)^{1 \over 1-p_c}\;.
\ee
 With these ingredients we can compute the volume
  of the $\epsilon$-box as
  \be\label{vole}
  {\rm Vol}\left[B^- ({\cal S}^*_\epsilon)\right] = \int_0^{\tau_\epsilon} d\tau \,(H\tau) \, \epsilon^d = {1\over 2H} \,(\tau_\epsilon H)^2 \,\epsilon^d\;,
  \ee
  and verify (\ref{epd}) with $a = 2(1-p_c)^{-1}$. 
  
For   solutions with matter degrees of freedom we need to check that the Lagrangian ${\cal L}_m$ is not too singular. For instance, if we have a  FRW terminal with metric 
\be\label{frw}
ds^2 = -d\tau^2 + a(\tau)^2 \,d\Sigma^2\;
\ee
and squared speed of sound $w=(2-\gamma)/\gamma$, the energy density scales as
\be
\rho \sim a(\tau)^{-d(1+w)} \sim \tau^{-2}\;.
\ee
Approximating the action dimensionally as the volume integral of the energy density, we estimate
\be\label{aces}
I\left[B^- ({\cal S}^*_\epsilon)\right]_{\rm bulk} \propto \epsilon^{\,d} \int_0^{\tau_\epsilon} d\tau \,\tau^\gamma {1\over \tau^2} \sim \epsilon^d \,\tau_\epsilon^{\gamma-1}\;.
\ee

The condition for the coarse-graining procedure to be well-defined is now
\be\label{genco}
{I\left[B^- ({\cal S}^*_\epsilon)\right]_{\rm bulk} \over {\rm Vol}_c [{\cal S}^*_\epsilon]} \sim \epsilon^{\,a}\;, \;\;\;a>0\;.
\ee
The physical condition that  the matter equation of state remains  strictly below the stiff limit, $w<1$, implies that $\gamma >1$ and thus (\ref{genco}) is satisfied 
 provided $\tau_\epsilon$ scales with a positive power of $\epsilon$. This
happens for any solution which {\it decelerates} away from the singularity, since the FRW conformal time is given by 
\be\label{frwco}
H\eta = \dfrac{d}{d-\gamma}\left(H \tau\right)^{d-\gamma \over d}\;.
\ee
It is precisely for decelerating singularities that we have $\gamma < d$ and $\tau_\epsilon \sim \epsilon^{d\over d-\gamma}$
scaling with a positive power of $\epsilon$, leading to an automatically well-defined coarse-graining limit.

\begin{figure}[t]
$$\includegraphics[width=9cm]{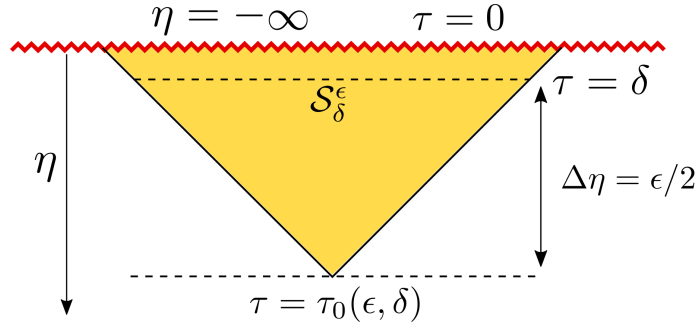} $$   
\begin{center}
\caption{\emph{The regularized terminal ${\cal S}^\delta_\epsilon$ and its past domain of dependence.  }\label{fig:frwacc}}
 \end{center}
\end{figure}

The situation is less clear for FRW metrics that {\it accelerate} away from the singularity, corresponding to $\gamma \geq d$. Now the FRW conformal time plummets to $-\infty$ as $\tau\rightarrow 0^+$. The problem in this case is that $D^- ({\cal S}^*_\epsilon)$ is not itself well defined, as any past light cone emanating from $\tau =0$ and converging at a finite value $\tau_0$ subtends an infinite comoving volume at the terminal surface. To address this point we regularize the terminal surface by bringing it slightly before the singularity at $\tau =\delta$, as indicated in Figure (\ref{fig:frwacc}).  In other words, we compute the past domain of dependence for a small, $\epsilon$-sized subset of $\Sigma_\delta$ rather than ${\cal S}^*$. Let us denote this set ${\cal S}^\delta_\epsilon$. Its past  domain of dependence, $D^- ({\cal S}^\delta_\epsilon)$, has an earliest proper time which is a function of both $\epsilon$ and $\delta$, 
\be
\tau_0 (\epsilon,\delta)  = {1\over H} \left((H\delta)^{d-\gamma \over d} + {d-\gamma \over d} {H\epsilon \over 2}\right)^{d \over d-\gamma}\;.
\ee
For $\gamma >d$, this quantity  vanishes linearly in $\delta$ as the terminal time cutoff is removed  at fixed $\epsilon$. 
Hence, when we repeat the estimate 
(\ref{aces}) we find that 
\be\label{newaces}
I\left[B^-({\cal S}^\delta_\epsilon)\right]_{\rm bulk} \propto \epsilon^{\,d} \int_{\delta}^{\tau_0} d\tau\, \tau^\gamma \,{1\over \tau^2} \sim \epsilon^d (\tau_0^{\,\gamma -1} - \delta^{\,\gamma-1}) \longrightarrow 0\;,
\ee
as $\delta \rightarrow 0$ at fixed $\epsilon$, since both terms vanish in the limit. Therefore, the bulk contribution vanishes when we remove the regularization at fixed comoving volume, even before we take $\epsilon \rightarrow 0$. 

The borderline case of a fluid with a stiff matter equation of state, i.e. $\gamma=1$, requires a separate analysis. From \eqref{aces} we can see that the bulk action diverges logarithmically near the singularity, so that a regularization procedure will be needed as well. Following the same notation as in Figure \ref{fig:frwacc}, we get

\begin{equation}
\label{logcondition}
{I\left[B^- ({\cal S}^\delta_\epsilon)\right]_{\rm bulk} \over {\rm Vol}_c [{\cal S}^\delta_\epsilon]} \sim \log\left(\dfrac{\tau_0(\epsilon, \delta)}{\delta} \right),
\end{equation}
whose behavior as  $\epsilon, \delta \rightarrow 0$ does depend on the order of the limits.
For fixed regularization parameter $\delta$ we have the expansion 

\begin{equation}
\dfrac{\tau_0(\epsilon, \delta)}{\delta} \sim 1+ \mathcal{O}(\epsilon),
\end{equation}
which makes \eqref{logcondition} approach linearly to zero as $\epsilon \rightarrow 0$. On the other hand, if we try to
remove the regularization at fixed $\epsilon$ we find 

\begin{equation}
\dfrac{\tau_0(\epsilon, \delta)}{\delta} \sim \dfrac{1}{H \delta} (H\epsilon)^{\frac{d}{d-1}}+ \mathcal{O}(\delta^{-\frac{1}{d}}),
\end{equation}
which diverges as $\delta \rightarrow 0$. Hence, the FRW singularity with stiff matter does not have a consistent coarse-graining limit
which would isolate the local complexity density.\footnote{The consistency of the  coarse-graining procedure would need to be analyzed anew if extra boundary terms  are included in the action,  beyond our minimal choice in (\ref{coma}) and (\ref{toal}).}

On the other hand,  the more `physical'  FRW singularities with $\gamma >1$  are extremely `thin' in any measure of local complexity. For $1<\gamma <d$, corresponding
to metrics that decelerate away from the singularity, we have a vanishing local contribution from the YGH term and a well-defined
coarse-graining procedure isolating this term. For $\gamma \geq d$, the situation is even more extreme, since  the quasilocal
complexity vanishes for sets of small but fixed comoving volume. 

We interpret these features as holographic analogs of the Weyl curvature criterion by Penrose, which also gave a smallest measure of complexity to FRW cosmologies. The main difference here is that the holographic notion of complexity is rather more refined, making a quantitative distinction between accelerating and decelerating cosmologies.

\section{Terminal Monotonicity}
\label{sec:mono}

\noindent

In this section we return to the full quasilocal complexity and discuss the monotonicity of its `nesting' properties.   Given that our definition of $C^*_u$  in (\ref{wdwu}) was  tailored to mimic the role of the cut surface ${\cal S}'_t$ in the  VC ansatz, we expect that $C^*_u$ should increase monotonically, at least asymptotically as $u\rightarrow u_*$. Since this may depend on the choice of `entropic' coupling $\lambda$, we conjecture that an appropriate choice exists such that this always happens. In other words,  we conjecture that
the full on-shell action of the WdW patch $I[{\cal W}_u]$  is either monotonically increasing as $u\rightarrow u^*$ or, in case it decreases, it does so at a rate bounded (in absolute magnitude) by  that of ${\rm Area}[{\cal V}_u]$. 

A priori, it is not immediately obvious that such monotonicity will hold. Let us consider again vacuum solutions satisfying (\ref{eme}). 
  Given a set of WdW patches ${\cal W}_u$ associated to a family of singular subsets ${\cal S}^*_u$ with $\gamma=1$, and holding
 codimension-two `edges' ${\cal V}_u$, we can write the full `nesting' complexity as a function of $u$ as arising from three contributions:
 \be\label{vacsplit}
 C^*_u \big |_{\rm VAC} = \alpha\,{\Lambda \over 4\pi G (d-1)} {\rm Vol}[{\cal W}_u] + \alpha \,{H \over 8\pi G} {\rm Vol}_c [{\cal S}^*_u] + {\lambda \over 4G} {\rm Area} [{\cal V}_u]\;.
 \ee
 The first term comes from the bulk Einstein--Hilbert action and is extensive in space-time volume of the WdW patch. The second term is the terminal density, extensive in the comoving volume of the terminal set, and the last term is the `entropic' counterterm. 

A glance at the expression (\ref{vacsplit}) indicates that the monotonicity is not obvious, in particular for
singularities embedded in AdS, since there one finds $\Lambda <0$ and the bulk contribution is negative as the space-time volume of the WdW patches grows. In addition, the particular monotonicity properties of ${\rm Area}[{\cal V}_u]$ could affect the final answer.  On the other hand, the space-time volume vanishes near  singularities of Einstein's equations.  Thus, we expect the main contribution to the 
on-shell action to come from the `corners' of the WdW patch, i.e. the vicinity of the sets ${\cal V}_u$. In such a situation we may expect that any threat to monotonicity coming from the bulk action could eventually be fixed by an appropriate choice of the entropic coupling $\lambda$. 

Short of a general proof, we have
examined a number of examples in which the explicit computation can be carried reliably and found agreement with the monotonicity conjecture. 
We consider here a few extreme cases  which illustrate the qualitatively different roles played by the entropic counterterm proportional to $\lambda$ in (\ref{defc}). Our choices are motivated by the ability to compute exactly the complexity on terminal WdW patches using the specific formula (\ref{vacsplit}),  but also by our interest in exposing as much as possible the contrast between entropy and complexity when referred to cosmological singularities. Quite generally, we can associate an entropic measure to a terminal set ${\cal S}^*$ by the limit of the codimension-two areas ${\rm Area}[{\cal V}_u]$ in Planck units, as $u\rightarrow u_*$, i.e.
\be\label{sen}
S[{\cal S}^*] \equiv \lim_{u\to u_*} {{\rm Area}[{\cal V}_u] \over 4G} 
\;.
\ee
With this definition, the full singularity of a standard black hole solution has finite entropy, whereas  any proper portion of the singularity has zero entropy. On the other hand, there are prototypical cosmological singularities with infinite entropy, such as the interior of a Coleman--de Luccia bubble in vacuum decay. It is very interesting to keep track of the terminal complexity in these wildly different situations from the point of view of the entropy as a measure of the `holographic dimensionality' of the relevant state spaces. 

Before embarking in our tour of examples, we would like to comment briefly on the relation to previous work. In \cite{BarbonRabinoComplexity}  the VC complexity was estimated for a number of cosmological singularities which are naturally embedded into concrete AdS/CFT constructions. In these examples it was found that a regularized version of the VC complexity was monotonically {\it decreasing} on approaching the singularity, in contrast with our statement here for the cuasilocal complexity. A similar behavior was obtained for the AC ansatz in the same examples by \cite{rabinonew}. The reason for this apparent discrepancy is simply that the full complexity computed in \cite{BarbonRabinoComplexity} is dominated by UV contributions to the VC ansatz, and these are highly dependent on the particular details of the embedding into asymptotically AdS geometries. For instance, some of the examples are based on singular CFT metrics which shrink to zero size, and others involve expanding domain walls in the bulk. In the first case it is natural that the UV contribution to complexity should have a negative derivative in time, as corresponds to a shrinking Hilbert space on the full CFT. In the second case, a time-dependent conversion between UV and IR degrees of freedom is introduced in the CFT by switching on a relevant operator with a time-dependent coupling, and the c-theorem explains why the UV again dominates the balance.    Therefore, there is no contradiction since the two monotonicity statements refer to different quantities. The positive monotonicity of the quasilocal complexity defined here (by restriction of the AC/VC ansatz to the interior of $D^- ({\cal S}^*)$,) is compatible with the negative monotonicity of the full complexity, particularly when  the latter is dominated by a strong UV time-dependence.

\subsubsection*{Vacuum terminals with constant entropy}

\noindent

We begin our tour of examples with the benchmark case of a   (future) black hole interior.
The standard case is provided by the  spherical AdS black hole solution, with an $\mathbb{R}\times SO(d)$ isometry group and metric 
\be\label{bhs}
ds^2 = -f(r)\,dt^2 + {dr^2 \over f(r)} + r^2 \,d\Omega_{d-1}^ 2\;, \qquad f(r) = 1+ {r^2 \over \ell^2} - {\mu \over r^{d-2}}\;,
\ee
where $\ell$ is the curvature radius of AdS. The vicinity of the singularity at $r=0$  is controlled by a single length scale, $\mu^{1\over d-2}$, which relates to the horizon radius $R$ through 
\be
\mu = R^{d-2} + {R^d \over \ell^2}\;.
\ee
 The near-terminal metric can be written in the form (\ref{locs}) with 
$\gamma =1 $ and
\be
H= \left({2 \over d \cdot \mu}\right)^{1\over d-2}\;.
\ee

The terminal set at  $r=0$ has topology ${\bf R} \times {\bf S}^{d-1}$, and its comoving metric degenerates through a stretching of the ${\bf R}$ factor and
a contraction of the sphere.  We can parametrize the terminal set by the homogeneous $t$-coordinate  along the ${\bf R}$ factor. Let ${\CS}^*_{\Delta t}$ denote a subset ${\bf I}_{\Delta t} \times {\bf S}^{d-1}$ of the terminal set, where ${\bf I}_{\Delta t} \subset {\bf R}$ is an interval of length $\Delta t$ along the $t$ coordinate. Its  comoving volume  is then given by 
\be
{\rm Vol}_c [{\CS}^*_{\Delta t}] = \Delta t \, \Omega_{d-1}\, H^{\,1-d}\;,
\ee
where $\Omega_{d-1} = {\rm Vol} [{\bf S}^{d-1}]$. 
The YGH contribution to the on-shell action  is 
\be\label{locc}
I[{\CS}^*_{\Delta t}]_{\rm YGH} = {H \over 8\pi G} \,{\rm Vol}_c [{\CS}^*_{\Delta t}]  = {d \,\mu  \,\Omega_{d-1} \,
\Delta t \over 16\pi G} 
 \;,
\ee
for both small and large black holes. Expressed in terms of physical quantities, we find a contribution to complexity which is positive
and proportional to   $\Delta t\,M$, where $M$ is the mass of the black hole. 

\begin{figure}[t]
$$\includegraphics[width=7cm]{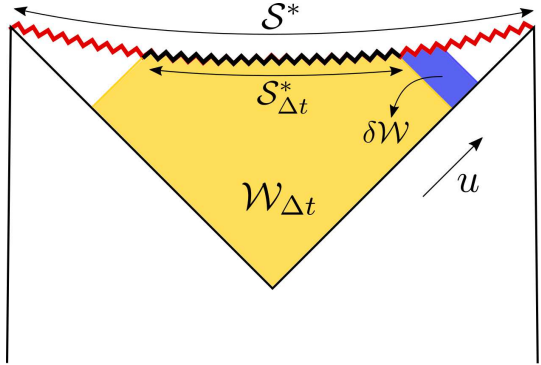} $$   
\begin{center}
\caption{\emph{ The terminal WdW patch ${\cal W}_{\Delta t}$ within the black-hole interior and the small stripe $\delta {\cal W}$. }\label{fig:bh}}
\end{center}

\end{figure}

The bulk contribution (\ref{bulkc}) is negative-definite. In order to compute it, we follow \cite{Poisson} 
and introduce  an infalling Eddington--Finkelstein coordinate
\be
u= t + \int^r {dr' \over f(r')}\;,
\ee
to write the metric in the form 
\be
ds^2 = -f(r)\,du^2 + 2 du dr + r^2 d\Omega_{d-1}^2\;.
\ee
This allows us to   compute the integral (\ref{bulkc})  as 
\be\label{bubu}
I[W_{\Delta t}]_{\rm bulk}  = -{ \Omega_{d-1} d \over 8\pi G \ell^2}   \int_{{\cal W}_{\Delta t}} du \,dr\, r^{d-1} 
\;.
\ee
To show monotonicity, it suffices to evaluate the action over a stripe  of $u$-extent $\delta u $, as indicated in Figure \ref{fig:bh}, and we obtain  
\be\label{resu}
I[\delta {\cal W}]_{\rm bulk}  = -{ \Omega_{d-1} d \over 8\pi G \ell^2}   \int_{\delta u} du \,\int_0^R dr\, r^{d-1}  =  -{\Omega_{d-1} \,R^d \over 8\pi G \ell^2} \,\delta u\;.
\ee
Adding the contribution from the YGH term and using   the relation
$
\mu = R^{d-2} + {R^d / \ell^2} 
$, 
we find  that 
\be
\delta I = \delta u {\Omega_{d-1} \,R^{d-2} \over 8\pi G \ell^2 } \left(  d \,\ell^2 + (d-1) R^2 \right)
\ee
is always positive, implying  the positive monotonicity of $C^*_{\Delta t}$. 
In establishing this result, the `entropic' counterterm   proportional to $\lambda$ plays no role,  since the horizon area  is asymptotically constant. 

\subsubsection*{Vacuum terminals with divergent entropy}

\noindent

A qualitatively different situation is obtained for terminals that look like standard cosmologies of FRW type. In general, such FRW models require non-trivial contributions from matter degrees of freedom. In order to stay within the realm of vacuum solutions, so that we can still apply (\ref{vacsplit}), we look at FRW metrics arising in the interior of vacuum bubbles of Coleman--de Luccia type.  A crucial property of any such bubble is that it expands, asymptotically approaching the speed of light, so that the area of bounding codimension-two surfaces is guaranteed to diverge, i.e. they always have divergent entropy. 

As an explicit example with an exact solution we can consider a kind of `topological crunch' spacetime (see for example \cite{chilenos, Banados, maldapim, BarbonRabinoComplexity}.)
   In this construction, we have a sort of higher-dimensional generalization of the  BTZ black hole, i.e. the metric is locally pure AdS$_{d+2}$, but a convenient identification by the group of integers realizes a time-dependent compactification with topology ${\bf S}^1 \times {\rm AdS}_{d+1}$, where the ${\bf S}^1$ fiber shrinks to zero size at the singular locus. More precisely, we consider the following  metric on $D^- ({\cal S}^*)$, 
 \be\label{topc}
 ds^2 = -dt^2 + \ell^2 \sin^2 (t/\ell) d{\bf H}_d^2 + \ell^2 \cos^2 (t/\ell) d\phi^2
 \;,
 \ee
 where $\phi$ is an angle parametrizing the additional compact circle, $d{\bf H}_d^2$ is the standard unit metric on the $d$-dimensional Euclidean hyperboloid and $\ell$ is the AdS radius of curvature. The singularity  at $t= \ell \pi/2$, occurring when  the compact circle degenerates to vanishing size, is very mild, but enough to guarantee that the terminal set ${\cal S}^*$ of topology ${\bf S}^1 \times {\bf H}_d$ has zero physical volume. The comoving volume of ${\cal S}^*$ is given by the standard volume forms on ${\bf S}^1 \times {\bf H}_d$ times $\ell^{d+1}$. 
 
 This model has the nice feature of admitting a straightforward embedding into AdS/CFT. On the exterior of $D^- ({\cal S}^*)$ we can realize the $SO(1,d)$ isometry on timelike de Sitter hypersurfaces instead of spacelike hyperbolic ones. Hence, the model can be embedded as a de Sitter-invariant state of a CFT$_{d+1}$ defined on dS$_d \times {\bf S}^1$. 
 
 \begin{figure}[t]
$$\includegraphics[width=8cm]{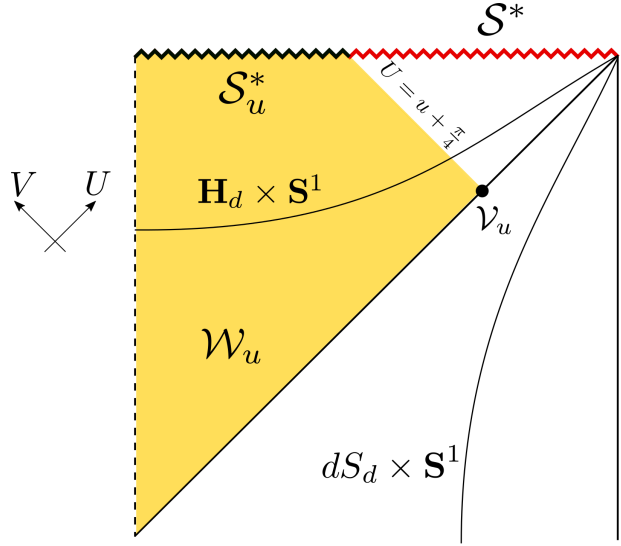} $$   
\begin{center}
\caption{\emph{ The causal structure of the topological crunch singularity and its associated WdW patches. The projection  of ${\cal W}_u$ onto the $(U,V)$ plane is shown as the yellow sector in the figure, and coincides with the domain of integration ${\cal W}(U,V)$. }\label{fig:tc}}
 \end{center}
\end{figure}

 In order to perform the required computations, it is useful to define null coordinates on $D^- ({\cal S}^*)$ as follows. First, we introduce a radial coordinate $\chi$ over ${\bf H}_d$ and a conformal time variable $\eta =2\tan^{-1} \left( e^{t/\ell}\right)$ over the AdS$_{d+1}$ factor, so that the metric is written in the form
 \be\label{mc}
 ds^2 =\ell^2 \sin^2 (t/\ell) \left[ -d\eta^2 + d\chi^2 + \sinh^2 (\chi) \,d\Omega_{d-1}^2\right] + \ell^2 \cos^2 (t/\ell) d\phi^2\;.
\ee
Next, we  introduce null coordinates  
\be\label{nulco}
\tan U=  e^{\eta + \chi} \;, \qquad \tan V = e^{\eta - \chi}
\;,
\ee
and write the complete metric as 
 \be\label{topcru}
 ds^2 = \ell^2 \sec^2 (U-V) \left[-4dUdV + \sin^2 (U-V) d\Omega_{d-1}^2  + \cos^2 (U+V) d\phi^2 \right]\;.
 \ee
 We consider a set of nested $(d+2)-$dimensional  WdW patches ${\cal W}_u$ bounded by the coordinate $u$, where $u = U -\pi / 4 \geq 0$. The bulk action is then given by
\be\label{bu}
I[{\cal W}_u]_{\rm bulk} = -{(d+1) \Omega_{d-1} \ell^d \over 2 G} \int_{{\cal W}(U,V)} dU dV {\tan^{d-1} (U-V)  \over \cos^3 (U-V)} \cos (U+V)\;,
\ee
where the domain of integration ${\cal W}(U,V)$ in the $(U,V)$ plane is shown in the Figure \ref{fig:tc}.  Notice that  Newton's constant $G$ has now the appropriate dimensionality for a $(d+2)-$dimensional spacetime, explaining the power of $\ell$ in the numerator.   A rather explicit expression can be obtained for the variation with respect to the $u$ coordinate,
\be\label{vbu}
{d \over du} I[{\cal W}_u]_{\rm bulk} = -{\Omega_{d-1}  \ell^d  \over 2 G d} \left[ (d-\sin (2u))( \tan (u+\pi/4))^d + \sin(2u) (\tan(2u))^d \right]\;,
\ee
 and the behavior near the singularity at $u_* = \pi/4$ is
 \be\label{nsing}
 {d \over du} I[{\cal W}_u]_{\rm bulk} \approx - {\Omega_{d-1}  \ell^d \over 2Gd} \left( d + 2^{-d} -1 \right) {1 \over (u_* -u)^d}
 \;.
 \ee
 A similarly explicit expression can be obtained for the variation of the YGH term:
 \be\label{vyg}
 {d \over du} I[{\cal S}^*_u]_{\rm YGH} = {\Omega_{d-1} \ell^d \over 2G} \,{(\tan (2u))^d \over \sin (2u)} \approx {\Omega_{d-1} \ell^d \over 2G}\,2^{-d} \,{1\over (u_* - u)^d}\;,
 \ee
 where the last expression is the asymptotic limit near the singularity. We find that, for all $d\geq 2$, the sum of (\ref{nsing}) and (\ref{vyg}) is negative, approaching minus-infinity as $u\rightarrow u_*$. On the other hand, one finds that the entropic term has the same degree of divergence near the singularity:
 \be\label{ent}
 {\rm Area}[{\cal V}_u] = 2\pi \,\Omega_{d-1} \ell^d \,(\tan(u+\pi/4))^{d-1} \;,
 \ee
 and the rate near the singuarity is given by 
 \be\label{ratent}
 {d \over du} {\rm Area}[{\cal V}_u] \approx 2\pi  \,\Omega_{d-1} \ell^d {d-1 \over (u_* - u)^d}\;.
 \ee
 
Hence, the terminal complexity of the topological crunch has the expected monotonicity property provided we choose
\be\label{tcbound}
{\lambda \over \alpha} > {1\over \pi d} \left(1-2^{-d}\right)\;.
\ee

This example differs  from the eternal black hole in that  the `entropy of the singularity' grows indefinitely, as measured by ${\rm Area}({\cal V}_u)$,  and the associated coupling $\lambda$ must satisfy a lower bound in order to guarantee monotonicity of $C^*_u$.

\begin{figure}[t]
$$\includegraphics[width=5cm]{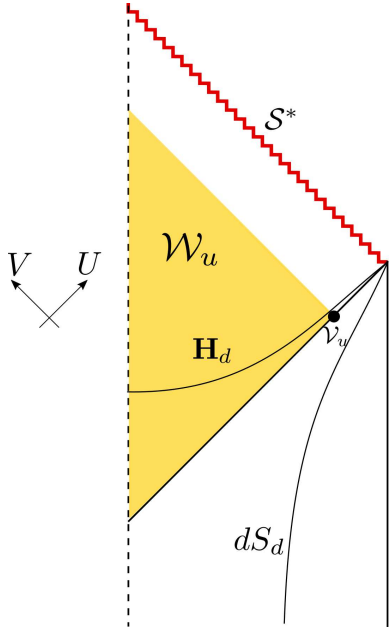} $$   
\begin{center}
\caption{\emph{ The idealized  state generating a null singularity by the collision of a thin-walled bubble with dS$_d$ worldvolume. }\label{fig:tw}}
 \end{center}
\end{figure}

An even simpler model of this kind is obtained by  removing the ${\bf S}^1$ factor in (\ref{topc}). The FRW patch of pure AdS$_{d+1}$ has a coordinate singularity at $t = \pi \ell$, but we may render it a true singularity by considering a thin-walled bubble with dS$_d$ wordvolume hitting the boundary of AdS$_{d+1}$ right at the boundary of the $t=\pi \ell$ null surface (cf. Figure \ref{fig:tw}). This model represents an approximate de Sitter-invariant condensate state for a CFT on dS$_d$ (cf. \cite{BarbonRabinoCrunches, BarbonRabinoconfcompl,  BarbonRabinoComplexity}.)

Using the same null coordinates introduced in (\ref{nulco}), we have a metric 
\be\label{frwk}
ds^2 = \ell^2 \sec^2 (U-V) \left[ -4 dU dV + \sin^2 (U-V) d\Omega_{d-1}^2 \right]
\;,\ee
The singularity sits at $U=\pi/2$ in the null limit. We can now consider the set of nested WdW patches ${\cal W}_u$ defined
by $0\leq U \leq u$ and $V\leq U$. These patches do not actually touch the singularity in the idealized null limit, as shown in Figure \ref{fig:tw},
so that the monotonicity properties of $C^*_u$ depend entirely on the balance between the negative-definite bulk contribution
and the volume of the codimension-two boundary sets ${\CV}_u$. The bulk action is given by
\be\label{bac}
I[{\cal W}_u]_{\rm bulk}= -{d \ell^{d-1} \Omega_{d-1} \over 8\pi G} \int_0^u dU \int_0^U dV \,{\tan^{d-1} (U-V) \over \cos^2 (U-V)} \approx -{\ell^{d-1} \Omega_{d-1} \over 8\pi G (d-1)} \left({1\over {\pi \over 2} - u}\right)^{d-1}\;,
\ee
where the last expression is an approximation for $u \approx \pi/2$. On the other hand, the volume of the codimension-two set is
\be\label{vvu}
{\rm Area} [{\cal V}_u] = \ell^{d-1} \Omega_{d-1} \,\tan^{d-1} (u) \approx  \ell^{d-1} \Omega_{d-1} \left({1\over {\pi \over 2} -u}\right)^{d-1}\;.
\ee
We thus conclude that monotonicity is guaranteed provided we pick an entropy coupling $\lambda$ satisfying the inequality
\be\label{monoc}
{\lambda \over \alpha} > {1 \over 2\pi (d-1)}\;.
\ee

In a sense, the singularity induced by a vacuum bubble in the thin-wall limit provides the extreme case in which the monotonicity
is threatened by a negative, diverging, bulk contribution. Here we see that this negative infinity is always tamed by a sufficiently large, but ultimately finite, entropic counterterm. It would be interesting to obtain these results in more realistic constructions going beyond the thin-wall approximation by the inclusion of explicit dynamical scalar fields.

\subsubsection*{A vacuum terminal with vanishing entropy}

\noindent

In this section we consider an example in which the entropy of the singularity, as defined by the volume of codimension-two sets ${\cal V}_u$, has precisely the opposite behavior to the topological crunch model,  namely it vanishes at the singular locus. Consider the  Kasner metric in $d+1$ dimensions 
\be\label{kasner}
ds^2 = -dt^2 + \sum_{i=1}^d (t H)^{2p_i} \,dx_i^2\;,
\ee
the standard vacuum solution with zero cosmological constant and $\mathbb{R}^{d}$ symmetry and, as discussed at length in the previous section,   a local approximation for `small portions' of more general singularities (cf. \cite{BKL, BKL2,BKL3,libro}.)
  We recall that the coefficients $p_i$ are restricted to satisfy $\sum_i p_i = \sum_i p_i^2 =1$, and  at least one of the exponents $p_i$ must be negative, indicating that at least one direction stretches as one approaches the singularity. The volume of codimension-two surfaces sitting at some constant value of $t$  always vanishes as $|t|$ in the  $t\rightarrow 0$ limit. 

 \begin{figure}[t]
$$\includegraphics[width=9cm]{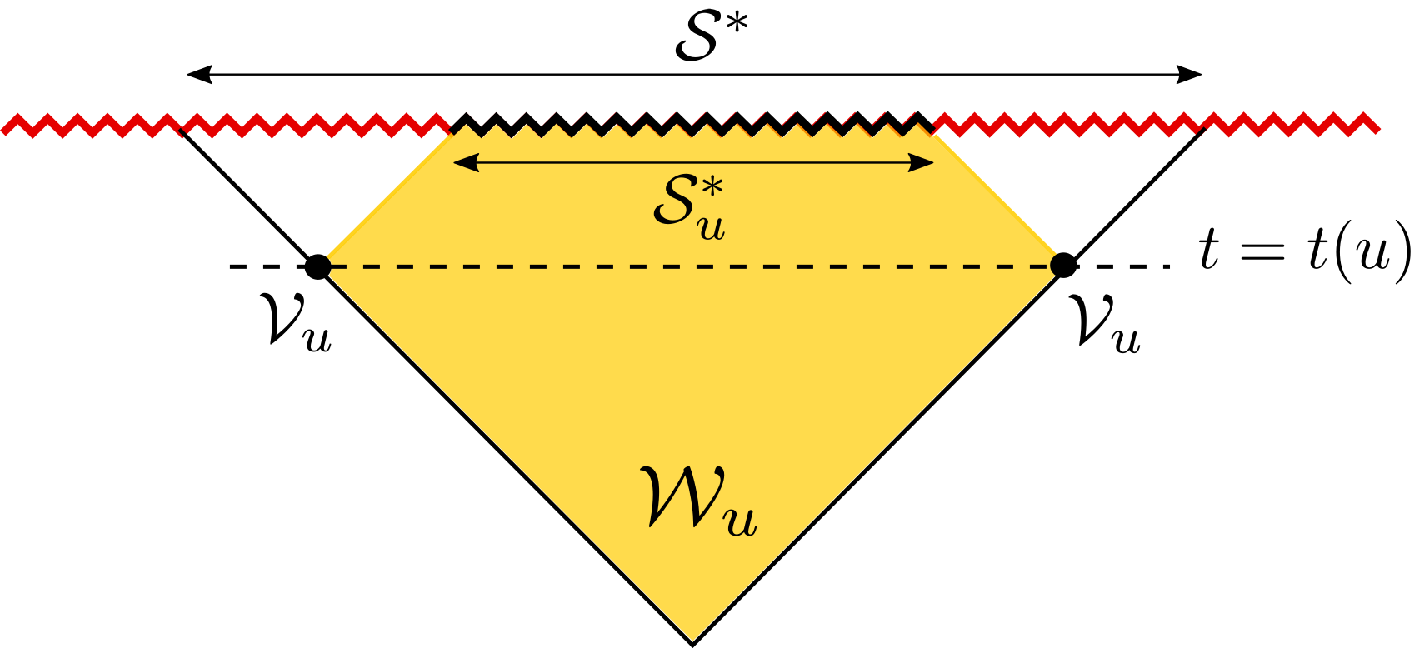} $$   
\begin{center}
\caption{\emph{ WdW patch for a Kasner slab ${\cal S}^*_u = \left[-{u \over 2} , {u\over 2} \right] \times {\bf R}^{d-1}$. }\label{fig:kasner}}
 \end{center}
\end{figure}

For simplicity of computations, we shall  consider a  family of sets ${\cal S}^*_u$  adapted to the symmetry of the metric, defined  as follows: in the ${\bf R}^{d}$ parametrized by the coordinates $x_j$, we single out one particular direction, $x_u$, and pick an interval  ${\bf I}_u =\left[-{u \over 2}, {u\over 2}\right]$ of length $u$ in this direction. The set of nested terminal sets is then defined as the `slabs' $ {\cal S}^*_u = {\bf R}^{d-1} \times{\bf  I}_u$. The full terminal set is obtained by taking the limit $u\rightarrow u_*$ with some finite $u_*$. The causal structure  of the WdW patches is shown in Figure \ref{fig:kasner}. The codimension-two surfaces ${\cal V}_u$ controlling the entropic coupling sit on the surfaces $t=t(u)$, with $t(u)$ determined by the equation 
\be
(1-p_u) (u_* - u) = 4 \,t(u)^{1-p_u}\;,
\ee
and $p_u$ is the Kasner exponent along the compact direction of the slab ${\cal S}^*_u$.  

 Since (\ref{kasner}) has already the form (\ref{locs}) with $\gamma =1$, we obtain an extensive and positive contribution from the YGH term 
\be\label{kygh}
I[{\cal S}^*_u]_{\rm YGH} = {H \over 8\pi G} \,{\rm Vol} [{\bf R}^{d-1}] \,u\;,
\ee
where we interpret the volume of ${\bf R}^{d-1}$ in the sense of defining the appropriate density along the non-compact directions. Moreover, the Ricci-flatness property of (\ref{kasner}) implies that there is no bulk contribution to the on-shell action, and we seem to obtain a monotonic result for $C^*_u$ with no explicit reference to the entropic coupling $\lambda$. 
On the other hand, a simple computation shows that 
\be\label{volen}
{\rm Area}[{\cal V}_u] = (1-p_u) H\,{\rm Vol} [{\bf R}^{d-1}] {u_* -u \over 2}\;,
\ee
which vanishes as $u\rightarrow u_*$, giving a monotonically {\it decreasing} contribution to complexity as soon as we have $\lambda >0$. Therefore, in order to ensure the right monotonicity property of $C^*_u$ we must assume that the entropic coupling satisfies an {\it upper} bound 
\be\label{ubound}
{\lambda \over \alpha} < {1 \over \pi (1-p_u)}
\;,
\ee
independently of whether the slab's finite interval is laid along a ripping ($p_u < 0$) or crunching ($p_u >0$) direction. 

Unlike the Coleman-de Luccia bubble, the Kasner situation has a finite complexity in the asymptotic limit $u\rightarrow u_*$, but
the entropy counterterm could make the total complexity approach the limit from above rather than from below. If we want to ensure a positive rate in this case we must prevent the entropic coupling from being too large, as indicated in 
the bound (\ref{ubound}).

\section{Conclusions and outlook}
\label{sec:conclusions}

\noindent

In this paper we have introduced quasilocal notions of AC complexity for terminal sets such as spacelike singularities in General Relativity. The basic idea is to build WdW patches restricted to the past causal domain of the singular set. Holographic data are associated to codimension-two surfaces on corresponding horizons. In principle, we can assign a notion of entropy to a singularity by looking at the area of these codimension-two sets. 

One of the  main observations made in this note is that the on-shell action of the WdW patches should be defined as monotonically increasing under the nesting of WdW patches, independently of the behavior of the entropy. We have tested this idea on a number of exactly solvable examples comprising the three qualitative behaviors with respect to the entropy:  asymptotically constant entropy like in black-hole interiors, diverging entropy as in the interior of Coleman--de Luccia bubbles, and vanishing entropy as in the Kasner spacetime. In all cases one can define monotonic terminal AC complexities at the price of adjusting an additive contribution proportional to the entropy. It would be interesting to back these checks with actual proofs, under the assumption of appropriate energy conditions. 

The YGH term evaluated at the singularity plays a special role. First, we emphasize that it is to be taken seriously despite the fact that
it is a contribution to the low-energy effective action extrapolated beyond its natural realm of applicability. Second, we have shown that one may isolate it as the local contribution to the complexity under an extreme coarse-graining procedure. In this respect, the relation of the YGH term  to the full quasilocal  complexity  is analogous to the relation between the  classical thermodynamic entropy, obtained through coarse-graining, and the exact von Neumann entropy of a quantum many-body state.  

Another interesting property of the local complexity is its  vague resemblance to Penrose's Weyl curvature criterion. In particular, FRW singularities are singled out by having   {\it vanishing}  complexity density, according to this definition. The similarity is not perfect though, since one can identify various differences. First, the `lack of complexity' seems to be even more severe for FRW metrics which accelerate away from the singularity. Second, within the local description of generic singularities, as presented in the classic BKL analysis, our ansatz assigns  
a vanishing  complexity density to the formal infinite sequences of chaotic Kasner `epochs'. Since these chaotic structures are generic in the light of the BKL analysis, we would conclude that the  local complexity density of generic spacelike singularities is zero. In this sense, complexity would behave similarly to local gravitational energy in General Relativity: while global and quasilocal definitions of gravitational energy exist, any attempt at a fully local definition is doomed to failure because of the equivalence principle. We find the parallel with
complexity unveiled here quite intriguing. 

A question of potential  interest is the generalization of these concepts to asymptotically de Sitter spacetimes. This is natural since the asymptotic
future of de Sitter is also a `causal terminal' and one may consider the behavior of the quantities defined in this paper. In fact, a simple check shows that every term in the quasilocal complexity ansatz becomes infinite in this case. Both the YGH term evaluated at the asymptotic future and the bulk action diverge. It turns out that in this case the YGH term dominates and formally gives   an infinitely  {\it negative} 
complexity. It would be interesting to elucidate these questions further.

\section{Acknowledgments}
\label{ackn}

We would like to thank M. Alishahiha,  B. Freivogel, C. Gomez, J. Maldacena, R. Myers, E. Rabinovici, J. Martinez-Magan and K. Sousa for discussions on various aspects of computational complexity, as well as the participants of the `It for Qbit Workshop' at Bariloche and `Complexity Workshop 2018' at AEI-Postdam, where preliminary versions of this work were presented. This work is partially supported by the Spanish Research Agency (Agencia Estatal de Investigaci\'on) through the grants IFT Centro de Excelencia Severo Ochoa SEV-2016-0597 and FPA2015-65480-P. The work of J.M.G. is funded by 
Fundaci\'on La Caixa under ``La Caixa-Severo Ochoa'' international predoctoral grant. 


\newpage


\bibliographystyle{utphys.bst}
\bibliography{refs}{}

\providecommand{\href}[2]{#2}\begingroup\raggedright\begin{thebibliography}{10}

\bibitem{BarbonRabinoComplexity}
J.~L.~F. Barb\'on and E.~Rabinovici, ``{Holographic complexity and spacetime
  singularities},'' \href{http://dx.doi.org/10.1007/JHEP01(2016)084}{{\em JHEP}
  {\bfseries 01} (2016) 084},
\href{http://arxiv.org/abs/1509.09291}{{\ttfamily arXiv:1509.09291 [hep-th]}}.

\bibitem{rabinonew}
S.~Bolognesi, E.~Rabinovici, and S.~R. Roy, ``{On Some Universal Features of
  the Holographic Quantum Complexity of Bulk Singularities},''
\href{http://arxiv.org/abs/1802.02045}{{\ttfamily arXiv:1802.02045 [hep-th]}}.

\bibitem{SusskindEntnotEnough}
L.~Susskind, ``{Entanglement is not enough},''
  \href{http://dx.doi.org/10.1002/prop.201500095}{{\em Fortsch. Phys.}
  {\bfseries 64} (2016) 49--71},
\href{http://arxiv.org/abs/1411.0690}{{\ttfamily arXiv:1411.0690 [hep-th]}}.

\bibitem{SusskindCAcorto}
A.~R. Brown, D.~A. Roberts, L.~Susskind, B.~Swingle, and Y.~Zhao,
  ``{Holographic Complexity Equals Bulk Action?},''
  \href{http://dx.doi.org/10.1103/PhysRevLett.116.191301}{{\em Phys. Rev.
  Lett.} {\bfseries 116} no.~19, (2016) 191301},
\href{http://arxiv.org/abs/1509.07876}{{\ttfamily arXiv:1509.07876 [hep-th]}}.

\bibitem{BrownSusskindAction}
A.~R. Brown, D.~A. Roberts, L.~Susskind, B.~Swingle, and Y.~Zhao,
  ``{Complexity, action, and black holes},''
  \href{http://dx.doi.org/10.1103/PhysRevD.93.086006}{{\em Phys. Rev.}
  {\bfseries D93} no.~8, (2016) 086006},
\href{http://arxiv.org/abs/1512.04993}{{\ttfamily arXiv:1512.04993 [hep-th]}}.

\bibitem{misner}
C.~W. Misner, ``{Mixmaster universe},''
\href{http://dx.doi.org/10.1103/PhysRevLett.22.1071}{{\em Phys. Rev. Lett.}
  {\bfseries 22} (1969) 1071--1074}.

\bibitem{BKL}
V.~A. Belinsky, I.~M. Khalatnikov, and E.~M. Lifshitz, ``{Oscillatory approach
  to a singular point in the relativistic cosmology},''
\href{http://dx.doi.org/10.1080/00018737000101171}{{\em Adv. Phys.} {\bfseries
  19} (1970) 525--573}.

\bibitem{BKL2}
V.~a. Belinsky, I.~m. Khalatnikov, and E.~m. Lifshitz, ``{A General Solution of
  the Einstein Equations with a Time Singularity},''
\href{http://dx.doi.org/10.1080/00018738200101428}{{\em Adv. Phys.} {\bfseries
  31} (1982) 639--667}.

\bibitem{BKL3}
V.~A. {Belinskii}, E.~M. {Lifshitz}, and I.~M. {Khalatnikov}, {\em {30 Years of
  the Landau Institute - Selected Papers. "Construction of a General
  Cosmological Solution of the Einstein Equation with a Time Singularity"}},
  \href{http://dx.doi.org/10.1142/9789814317344_0077}{pp.~763--766}.
\newblock World Scientific Publishing Co, 1996.

\bibitem{libro}
V.~Belinski and M.~Henneaux, {\em {The Cosmological Singularity}}.
\newblock Cambridge University Press,
2017.
\newblock

\bibitem{Penrose}
R.~Penrose, ``{Singularities and time-asymmetry},'' in {\em {General
  Relativity: An Einstein Centenary Survey}}, pp.~581--638.
\newblock
1979.
\newblock

\bibitem{rt}
S.~Ryu and T.~Takayanagi, ``{Holographic derivation of entanglement entropy
  from AdS/CFT},'' \href{http://dx.doi.org/10.1103/PhysRevLett.96.181602}{{\em
  Phys. Rev. Lett.} {\bfseries 96} (2006) 181602},
\href{http://arxiv.org/abs/hep-th/0603001}{{\ttfamily arXiv:hep-th/0603001
  [hep-th]}}.

\bibitem{hrt}
V.~E. Hubeny, M.~Rangamani, and T.~Takayanagi, ``{A Covariant holographic
  entanglement entropy proposal},''
  \href{http://dx.doi.org/10.1088/1126-6708/2007/07/062}{{\em JHEP} {\bfseries
  07} (2007) 062},
\href{http://arxiv.org/abs/0705.0016}{{\ttfamily arXiv:0705.0016 [hep-th]}}.

\bibitem{Mohsen}
M.~Alishahiha, ``{Holographic Complexity},''
  \href{http://dx.doi.org/10.1103/PhysRevD.92.126009}{{\em Phys. Rev.}
  {\bfseries D92} no.~12, (2015) 126009},
\href{http://arxiv.org/abs/1509.06614}{{\ttfamily arXiv:1509.06614 [hep-th]}}.

\bibitem{lloyd}
S.~{Lloyd}, ``{Ultimate physical limits to computation},''
  \href{http://dx.doi.org/10.1038/35023282}{{\em Nature} {\bfseries 406} (Aug.,
  2000) }, \href{http://arxiv.org/abs/quant-ph/9908043}{{\ttfamily
  quant-ph/9908043}}.

\bibitem{montero}
W.~Cottrell and M.~Montero, ``{Complexity is simple!},''
  \href{http://dx.doi.org/10.1007/JHEP02(2018)039}{{\em JHEP} {\bfseries 02}
  (2018) 039},
\href{http://arxiv.org/abs/1710.01175}{{\ttfamily arXiv:1710.01175 [hep-th]}}.

\bibitem{Myerstdep}
D.~Carmi, S.~Chapman, H.~Marrochio, R.~C. Myers, and S.~Sugishita, ``{On the
  Time Dependence of Holographic Complexity},''
  \href{http://dx.doi.org/10.1007/JHEP11(2017)188}{{\em JHEP} {\bfseries 11}
  (2017) 188},
\href{http://arxiv.org/abs/1709.10184}{{\ttfamily arXiv:1709.10184 [hep-th]}}.

\bibitem{BarbonMartinHyperbolic}
J.~L.~F. Barb\'on and J.~Mart\'in-Garc\'ia, ``{Holographic Complexity Of Cold
  Hyperbolic Black Holes},''
  \href{http://dx.doi.org/10.1007/JHEP11(2015)181}{{\em JHEP} {\bfseries 11}
  (2015) 181},
\href{http://arxiv.org/abs/1510.00349}{{\ttfamily arXiv:1510.00349 [hep-th]}}.

\bibitem{MyersFormation}
S.~Chapman, H.~Marrochio, and R.~C. Myers, ``{Complexity of Formation in
  Holography},'' \href{http://dx.doi.org/10.1007/JHEP01(2017)062}{{\em JHEP}
  {\bfseries 01} (2017) 062},
\href{http://arxiv.org/abs/1610.08063}{{\ttfamily arXiv:1610.08063 [hep-th]}}.

\bibitem{noncomp}
J.~L.~F. Barb\'on and J.~Mart\'in-Garc\'ia, ``{Holographic non-computers},''
  \href{http://dx.doi.org/10.1007/JHEP02(2018)181}{{\em JHEP} {\bfseries 02}
  (2018) 181},
\href{http://arxiv.org/abs/1710.06415}{{\ttfamily arXiv:1710.06415 [hep-th]}}.

\bibitem{Poisson}
L.~Lehner, R.~C. Myers, E.~Poisson, and R.~D. Sorkin, ``{Gravitational action
  with null boundaries},''
  \href{http://dx.doi.org/10.1103/PhysRevD.94.084046}{{\em Phys. Rev.}
  {\bfseries D94} no.~8, (2016) 084046},
\href{http://arxiv.org/abs/1609.00207}{{\ttfamily arXiv:1609.00207 [hep-th]}}.

\bibitem{Ross}
A.~Reynolds and S.~F. Ross, ``{Divergences in Holographic Complexity},''
  \href{http://dx.doi.org/10.1088/1361-6382/aa6925}{{\em Class. Quant. Grav.}
  {\bfseries 34} no.~10, (2017) 105004},
\href{http://arxiv.org/abs/1612.05439}{{\ttfamily arXiv:1612.05439 [hep-th]}}.

\bibitem{chilenos}
M.~Ba\~nados, A.~Gomberoff, and C.~Martinez, ``{Anti-de Sitter space and black
  holes},'' \href{http://dx.doi.org/10.1088/0264-9381/15/11/018}{{\em Class.
  Quant. Grav.} {\bfseries 15} (1998) 3575--3598},
\href{http://arxiv.org/abs/hep-th/9805087}{{\ttfamily arXiv:hep-th/9805087
  [hep-th]}}.

\bibitem{Banados}
M.~Ba\~nados, ``{Constant curvature black holes},''
  \href{http://dx.doi.org/10.1103/PhysRevD.57.1068}{{\em Phys. Rev.} {\bfseries
  D57} (1998) 1068--1072},
\href{http://arxiv.org/abs/gr-qc/9703040}{{\ttfamily arXiv:gr-qc/9703040
  [gr-qc]}}.

\bibitem{maldapim}
J.~Maldacena and G.~L. Pimentel, ``{Entanglement entropy in de Sitter space},''
  \href{http://dx.doi.org/10.1007/JHEP02(2013)038}{{\em JHEP} {\bfseries 02}
  (2013) 038},
\href{http://arxiv.org/abs/1210.7244}{{\ttfamily arXiv:1210.7244 [hep-th]}}.

\bibitem{BarbonRabinoCrunches}
J.~L.~F. Barb\'on and E.~Rabinovici, ``{AdS Crunches, CFT Falls And
  Cosmological Complementarity},''
  \href{http://dx.doi.org/10.1007/JHEP04(2011)044}{{\em JHEP} {\bfseries 04}
  (2011) 044},
\href{http://arxiv.org/abs/1102.3015}{{\ttfamily arXiv:1102.3015 [hep-th]}}.

\bibitem{BarbonRabinoconfcompl}
J.~L.~F. Barb\'on and E.~Rabinovici, ``{Conformal Complementarity Maps},''
  \href{http://dx.doi.org/10.1007/JHEP12(2013)023}{{\em JHEP} {\bfseries 12}
  (2013) 023},
\href{http://arxiv.org/abs/1308.1921}{{\ttfamily arXiv:1308.1921 [hep-th]}}.

\end{thebibliography}\endgroup

%
%

\end{document}